\documentclass[12pt,preprint]{aastex}

\shorttitle{Characterizing the disk around 2M1207}
\shortauthors{Riaz \& Gizis}

\begin{document}

\title{Characterizing the disk around the TW Hydrae Association brown dwarf 2MASSW J1207334-393254}

\author{Basmah Riaz and John E. Gizis}
\affil{Department of Physics and Astronomy, University of Delaware,
    Newark, DE 19716; basmah@udel.edu, gizis@udel.edu}

\begin{abstract}

We present detailed modeling of the disk around the TW Hydrae Association (TWA) brown dwarf 2MASSW J1207334-393254 (2M1207), using {\it Spitzer} observations from 3.6 to 24 $\micron$. The spectral energy distribution (SED) does not show a high amount of flaring. We have obtained a good fit using a flat disk of mass between $10^{-4}$ and $10^{-6}$ $M_{\sun}$, $\dot{M}$ $\la10^{-11} M_{\sun}$ $yr^{-1}$ and a large inclination angle between 60$\degr$ and 70$\degr$. We have used three different grain models to fit the 10 $\micron$ Si emission feature, and have found the results to be consistent with ISM-like dust. In comparison with other TWA members, this suggests lesser dust processing for 2M1207 which could be explained by mechanisms such as aggregate fragmentation and/or turbulent mixing. We have found a good fit using an inner disk radius equal to the dust sublimation radius, which indicates the absence of an inner hole in the disk. This suggests the presence of a small K-$L^{\prime}$ excess, similar to the observed K-[3.6] excess.

\end{abstract}


\section{Introduction}

Over the last decade, a large number of sub-stellar mass objects have been discovered, with masses ranging from the hydrogen burning limit ($\la$ 0.075 $M_{\sun}$) down to the mass of giant planets and below the deuterium burning limit ($\la$ 0.013 $M_{\sun}$). The presence of accretion disks in sub-stellar objects has been confirmed by signatures of ongoing accretion (e.g. Muzerolle et al. 2003, 2005; Mohanty et al. 2005) and the presence of excess emission in the near- and mid-infrared (e.g. Jayawardhana et al. 2003; Muench et al. 2001; Mohanty et al. 2004; Apai et al. 2004; Luhman et al. 2005). The fraction of circumstellar disks appears to be similar for young brown dwarfs and T Tauri stars (e.g. Luhman et al. 2005; Jayawardhana et al. 2003). Furthermore, brown dwarf disks have been found to show a range in disk properties, similar to T Tauri disks. The observed SEDs have been well fit by models of both flared and flat disks using a range of grain sizes from ISM-like dust to grains as large as 1 mm (e.g. Allers et al. 2006; Muzerolle et al. 2006). Inner holes of a few sub-stellar radii have been found to be common in brown dwarf disks (e.g. Mohanty et al. 2004; Allers et al. 2006). Recently, Muzerolle et al. (2006) have reported the first brown dwarf with evidence of an AU-scale inner disk hole. The inner disk lifetimes for brown dwarf disks appear to be similar to the disks around T Tauri stars (Jayawardhana et al. 2003). While some brown dwarfs demonstrate an absence of the silicate emission feature near 10 $\micron$ (e.g. Cha H$\alpha$2; Apai et al. 2002), others display a variety in the shape and strength of this feature, that have been explained by the presence of either small ISM grains (e.g. Cha H$\alpha$1; Sterzik et al. 2004) or large amorphous grains of sizes $\sim$2 $\micron$ (e.g. CFHT BD Tau 4; Apai et al. 2004). Disk masses in the range -4 $\la$ log ($M_{d}/M_{\sun}$) $\la$ -7 have been determined for young brown dwarf disks (e.g. Walker et al. 2004), with no change seen in the relative disk mass with object mass at the substellar boundary (Scholz et al. 2006). A similar absence of discontinuity has been seen in the distribution of disk accretion rates with stellar mass (Muzerolle et al. 2005). Together such similarities indicate that brown dwarf disks are just scaled-down versions of T Tauri disks.

A detailed study of brown dwarf disks is important to test the different formation mechanisms for these sub-stellar objects. The numerical simulations of star formation by Bate et al. (2003) have shown that for brown dwarfs that form via the ejection mechanism, i.e., as stellar embryos in multiple systems that are ejected at an early stage, disks of radii larger than $\sim$20 AU are rare ($\sim$ 5\%). This is because brown dwarfs are ejected so soon after their formation that they do not have time to accrete the high angular momentum gas required to form a large disk. On the other hand, if most brown dwarf disks observed have large outer radii, it would suggest that the ejection mechanism rarely operates and would support a star-like formation, i.e. via the collapse of molecular cloud cores with sub-stellar masses. Planet-like formation of a brown dwarf in a circumstellar disk would, however, require an absence of dust around it (e.g. Apai et al. 2002). Recent efforts by Scholz et al. (2006) have shown that 25\% of their target brown dwarfs have disk radii $>$10 AU, indicating star-like formation. Observations of larger samples of sub-stellar disks would be able to confirm or reject the different formation scenarios. Characterization of brown dwarf disks is also important to understand the conditions under which planets can form in these disks. Processes that may lead to planet formation such as grain growth, crystallization and dust settling have been reported for some brown dwarfs (e.g. Apai et al. 2005), suggesting that even sub-stellar disks of a few Jupiter masses can harbor planets.

The brown dwarf 2MASSW J1207334-393254 (hereafter, 2M1207) was discovered by Gizis (2002) and was confirmed to be a TW Hydrae Association (TWA) member by Scholz, R. -D. et. al. (2005), based on its proper motion. It is known to be undergoing accretion that varies by a factor of 5-10 on a time scale of $\sim$6 weeks (Scholz et al. 2005). Excess emission has been detected in the infrared (IR) by Sterzik et al. (2004; hereafter S04) and Riaz et al. (2006; hereafter R06), indicating the presence of a circumsubstellar disk of gas and dust. The UV spectrum for 2M1207 shows $H_{2}$ emission lines as due to the presence of circumsubstellar gas, that gets (shock) heated up to high temperatures of $\sim10^{5}$ K as it falls along an accretion column onto the substellar surface, giving rise to C IV lines in emission (Gizis et al. 2005). Gizis et al. had also shown that the lack of Si emission in the hot spots is due to its depletion into dust grains in the circumsubstellar disk. Indeed, observations by S04 indicated the presence of Si emission in the dusty disk. Based on their 8.7 and 10.4 $\mu$m observations, S04 had found that both a large grain size (up to 5 $\micron$), large inclination angle flared disk, {\it and} a flat disk composed of small (0.1 $\micron$) or large (2 $\micron$) grains provide a good fit. In R06, we had reported on the {\it Spitzer} observations of 2M1207 from 3.6 to 24 $\micron$, and had shown the presence of warm (T$>$100 K) dust close (R$<$0.2 AU) to the brown dwarf. We were able to get a good fit to our observations using a straight line of slope of $\sim$ -1.3, indicating a flat disk for 2M1207. Here we have used disk models to further characterize the disk.

\section{Disk modeling}

We have used the 2-D radiative transfer code by Whitney et al. (2003). The circumstellar geometry consists of a rotationally flattened infalling envelope, bipolar cavities, and a flared accretion disk in hydrostatic equilibrium. With evolution from a Class 0 to Class III source, the cavity density and the envelope infall rate decreases, while the disk radius and the cavity opening angle increases. The disk density is proportional to $\varpi^{-\alpha}$, where $\varpi$ is the radial coordinate in the disk midplane, and $\alpha$ is the radial density exponent. The disk scale height increases with radius, $h=h_{0}(\varpi / R_{*})^{\beta}$, where $h_{0}$ is the scale height at $R_{*}$ and $\beta$ is the flaring power. Since we are fitting a disk source, the envelope was turned off by setting its mass infall rate equal to zero. For the stellar parameters, we have used $T_{eff}$=2550K and $M_{*}$=0.024 $M_{\sun}$ (Mohanty et al. 2006). Mohanty et al. have considered an age of 5-10 Myr to determine the mass. Using this $T_{eff}$ and $M_{*}$, the evolutionary tracks by Burrows et al. (1997) imply $R_{*} \sim$ 0.024 $R_{\sun}$. A distance of 59 pc (Song et al. 2006) was used to scale the output fluxes from the models to the luminosity and distance of 2M1207. The NextGen (Hauschildt et al. 1999) atmosphere file for a $T_{eff}$ of 2600K, and log {\it g} = 3.5 was used to fit the atmosphere spectrum of the central sub-stellar source. We note that Mohanty et al. have determined the effective temperature by using the DUSTY models by Allard et al. (2001). However, we could not find any differences between the two atmospheric models at a $T_{eff}$ of 2600 K. As noted by Gorlova et al. (2003), variations in these models due to the inclusion of dust opacities become apparent only below $\sim$2300 K. Table 1 lists the IRAC and MIPS observations from R06. Table 2 lists the stellar parameters used, and the range of disk parameters that provide good fits to 2M1207.

Fig. 1 compares the amount of flaring in 2M1207 disk with other brown dwarf disks, by looking at the ratio of 24 and 4.5 $\micron$ fluxes as a function of the 24 $\micron$ flux. The dashed line represents a geometrically thin, optically thick flat disk with a spectral slope $\lambda F_{\lambda} \propto \lambda^{-4/3}$. The solid line represents the photosphere of a brown dwarf at $T_{eff}$=2600K. L1707 and L291 are two brown dwarf disks in IC 348 that show some amount of flaring, and have been fit by a model disk with $\dot{M}$ = $10^{-11} M_{\sun}$ $yr^{-1}$, $a_{max}$ = 0.25 $\micron$ and an inclination of 60$\degr$ (Muzerolle et al. 2006). 2M1139511-315921 (hereafter, 2M1139), another brown dwarf in TWA, is nearly photospheric, while TW Hya has a highly flared disk. 2M1207 lies just at the dashed line, indicating an optically thick flat disk. That the disk is optically thick can also be inferred from the 3.6 to 8 $\micron$ slope, $\alpha$, of the SED. Lada et al. (2006) have used the values of $\alpha$ to discriminate between sources with and without disks in IC 348. Sources with $\alpha >$ -1.8 have optically thick disks, while the ones with -1.8 $> \alpha >$ -2.65 have anemic or optically thin disks. A source with $\alpha <$ -2.56 is photospheric (i.e. disk-less). An $\alpha$ of -1.5 for 2M1207 thus indicates the presence of an optically thick disk.

For a disk in vertical hydrostatic equilibrium, the amount of flaring in the disk depends on the stellar mass and the disk temperature, as the disk scale height {\it h} $\propto (T_{d}/M_{*})^{1/2}$ (Walker et al. 2004). Thus the scale height increases with decreasing mass, resulting in more vertically extended disks for brown dwarfs compared to those around classical T Tauri stars (CTTS). Walker et al. found {\it h}(100AU) for a 0.01 $M_{\sun}$ brown dwarf to be $\sim$60 AU, compared to $\sim$15 AU for a CTTS. Due to the larger scale heights for lower masses, the disk can intercept and thus scatter and reprocess more stellar radiation, resulting in large infrared excesses. We have varied the degree of flaring by adjusting the values of the flaring power $\beta$. Fig. 2a shows that for $M_{2M1207} \sim 0.024 M_{\sun}$, the SED is highly flared for $\beta$ = 1.25 (typical value for models of T Tauri disks in hydrostatic equilibrium). The variations in the SEDs as $\beta$ is lowered are more evident for $\lambda$ $>$ 10 $\micron$. Our 24 $\micron$ observation indicates that the models with $\beta$ between 1.05 and 1.1 are good fits to 2M1207 disk.  

The models we are using here consider a flared accretion disk. Walker et al. (2004) have shown that low mass flared disks are a good fit to SEDs that had been previously fit by a flat disk. Even though the 
hydrostatic solution gives a highly flared disk, most of the mass is in the gas due to a gas-to-dust
mass ratio of 100. So if there is dust settling in a flared disk, the dust disk observed in the IR could still be flattened compared to the gas disk. Walker et al. have concluded that long wavelength observations are required to discriminate between a flat disk and a low mass flared disk. The typical range of brown dwarf disk masses is found to be -4 $\la$ log ($M_{d}/M_{\sun}$) $\la$ -7 (e.g. Walker et al. 2004, Scholz et al. 2006). In Fig. 2b, we have varied the disk masses between $10^{-2} M_{\sun}$ and $10^{-9} M_{\sun}$. Higher mass disks are more flared, while the $10^{-9} M_{\sun}$ SED is nearly photospheric. As the disk mass decreases, the disk becomes optically thin, and near and mid-infrared radiation is seen from the whole disk, whereas for larger disk masses, the disk is optically thick and IR radiation is only seen from the surface layers of the disk. Variations in the SEDs are more evident for $\lambda$$>$10 $\micron$, and our 24 $\micron$ observation indicates disk masses between $10^{-4}$ and $10^{-6}$ $M_{\sun}$ to be good fits. Observations at far-IR/sub-mm wavelengths can correctly determine the disk mass and confirm the presence of a population of large grains that dominate the disk mass but contribute little to the opacity at near- and mid-IR wavelengths.

In Fig. 2c, we have explored a range of inclination angles to the line of sight. Due to binning of photons in the models, there are a total of 10 viewing angles, with face-on covering 0-18$\degr$ inclinations. The mid- and far-IR fluxes increase with decreasing inclinations, as the emission at these wavelengths is from an optically thick disk, while the optically thin millimeter fluxes are independent of the inclination.  For edge-on disks, the star contributes only indirectly through scattered light. The 10 $\micron$ silicate feature is a broad absorption band for such sources, since the large extinction in the disk midplane blocks thermal radiation from the deeper layers, and low albedos prevent radiation from scattering out through the upper disk layers (Whitney et al. 2003). Fig. 2c shows that bins of 63$\degr$ and 69$\degr$ are a good fit to 2M1207 disk. On the basis of the redshifted absorption component in the H$\alpha$ profile, Scholz et al. (2005) had concluded that 2M1207 is seen with an inclination of {\it i} $\ga$ 60$\degr$. Our model fits are thus consistent with previous conclusion on the inclination angle.

A well-mixed model assumes diffuse ISM-like dust (i.e. silicates and graphite grains with a maximum size of $\sim$0.25 $\micron$) to be uniformly mixed with the disk gas. However, such models exhibit too little mm-flux and larger far-infrared fluxes than those observed for CTTS (D$\arcmin$Alessio et al. 2006). Indeed,  Scholz et al. (2006) could not find a good fit to their Taurus brown dwarf disks using models with dust and gas well mixed in vertical hydrostatic equilibrium, as these produce too large far-infrared emission, while underpredict the mid-infrared fluxes. The flat SED slopes observed at mm-wavelengths in some T Tauri disks indicate the presence of grains larger than those in the ISM. This suggests dust settling and grain growth that affect disk temperatures and vertical structures, resulting in dust photospheres that are flatter rather than flared (e.g., Dullemond \& Dominik 2004). With grain evolution, the upper disk layers are rapidly depleted of the dust material, leaving only the smallest grains at larger scale heights while the larger grains settle in the disk midplane. With reduction in number of small grains in the upper layers, the IR opacity decreases, which results in lower IR fluxes. Also, this small population of small grains at larger scale heights produces the observed emission features such as the silicate bands near 10 $\micron$. The large grains that settle in the interior are able to produce large mm fluxes (due to their larger mm emissivity) and can explain the observed flat SED slopes at mm wavelengths (D$\arcmin$Alessio et al. 2006). Thus models in which grain sizes vary with the scale height in the disk are better able to fit the observed SEDs, compared to the well-mixed models. 

The models used here provide the capability to include different grains in different regions (see Whitney et al. 2003, Fig. 1). There are three grain models supplied by the code. All of these grain models use a range of grain sizes that can be approximated as a power-law. The difference is in the {\it maximum} grain sizes. The three grain models used are: large grains with a size distribution that decays exponentially for sizes larger than 50 $\micron$ up to 1 mm, grains of sizes with $a_{max}\sim$ 1 $\micron$, and ISM-like grains with $a_{max} \sim$ 0.25 $\micron$. We have varied these three different grain sizes in the disk midplane and the upper atmosphere. The upper atmosphere corresponds to the highest layers in the disk, while the disk midplane is the densest region. The code, by default, uses the large grains of $a_{max}$=1mm in the disk midplane, the $a_{max}\sim$1 $\micron$ grains in the disk atmosphere, and the ISM-like grains in the outflow. In short, the grain size decreases with increasing scale height in the disk, with the small grains in the upper layers and large grains in the midplane. 

Using these default parameters, while the model SED is a good fit to our IRAC and 24 $\micron$ observations (Fig. 3a), it fails to produce the silicate emission feature near 10 $\micron$, as can be seen from Fig. 3b. Wavelengths between 7 and 14 $\micron$ cover the 10 $\micron$ silicate emission feature, whose shape and strength is determined by dust grain size and composition. A peak at 9.8 $\micron$ indicates the presence of amorphous interstellar grains, while disks that show prominent crystalline silicate emission features, indicative of highly processed dust, have characteristic peaks at 9.3 and 11.3 $\micron$ (e.g., Apai et al. 2005). For the case of 2M1207, there are only three observations (at 8, 8.7 and 10.4 $\micron$) in this wavelength range to indicate the presence of a silicate emission feature. We have used a simple reduced $\chi^{2}$ analysis to determine if the large or the small grains provide a good fit to this feature. Fig. 3b shows that if the disk midplane grain model is changed from $a_{max}$=1mm to the $a_{max}\sim$1 $\micron$ sized grains, the SED is a good fit to the 8.7 $\micron$ observation, but the fluxes are still low to fit the 10.4 $\micron$ point. On the other hand, reducing the grain size further to ISM-type grains with $a_{max} \sim$ 0.25 $\micron$ provides the best fit (lowest reduced $\chi^{2}$ value) to the silicate feature. For our IRAC and 24 $\micron$ observations, any grain model provides a good fit, as variations in the SEDs for different grain sizes are evident at either $\sim$10 $\micron$ or at far-IR/sub-mm wavelengths. The presence of large grains in the disk midplane results in flatter slopes at longer wavelengths, as can be seen from Fig. 3a. 

The changes in the grain sizes discussed above were made in the disk midplane, while setting the disk atmosphere grains to sizes of $a_{max}\sim$1 $\micron$. We next varied the disk upper atmosphere grain sizes, while setting the disk midplane grains to $a_{max}\sim$1 mm. In the disk atmosphere, using ISM-like grains or any other grain size fails to produce the silicate emission feature. The 10 $\micron$ silicate feature only seems to be sensitive to the grains in the disk midplane. This can be explained by the difference in the $\tau$=1 surface depth as a function of wavelength of the emission. Though the 10 $\micron$ silicate feature arises from the optically thin surface layer of the disk, it probes a deeper layer compared to shorter wavelengths. The upper atmosphere corresponds to the highest layers of the disk, where the disk material is optically thick to $\lambda \sim$1 $\micron$ stellar radiation. At $\lambda \sim$10 $\micron$, characteristic of the reprocessed radiation from the top disk layers, the disk has less optical depth and the reprocessed radiation can diffuse and heat up the inner layers of the disk. The disk atmosphere is heated up to a higher temperature than the inner layers, resulting in a vertical temperature inversion that produces the silicate feature in emission (Calvet et al. 1992). The model used here assigns large grains to the high density regions. Due to this, when large grains are used in the disk midplane, similar grains are placed close to the inner wall since it is of high density. This affects the emission from the inner wall and hence the observed flux at 10 $\micron$. Thus by using sub-micron sized grains in the disk midplane, we have indirectly reduced the inner wall grain size, and were able to obtain a good fit to the 10 $\micron$ Si feature. There could still be bigger grains in the really dense midplane region of the disk, the presence of which can be confirmed with far-IR/sub-mm observations. On the other hand, if this were a $\arcsec$three-layered$\arcsec$ disk model instead of a $\arcsec$two-layered$\arcsec$ one, large grains could be placed in the very dense regions without affecting the 10 $\micron$ feature.

On the basis of a reduced $\chi^{2}$ analysis, we have found that ISM-like grains provide the best fit to the 10 $\micron$ silicate emission feature for 2M1207. This is in contrast with other TWA members, such as TW Hya, HD98800B and Hen 3-600A, that exhibit broad silicate emission features. The N-band spectra for Hen 3-600A shows a mixture of crystalline silicate components (Honda et al. 2003), while those for TW Hya and HD98800B have been fit using large (2 $\micron$) amorphous olivine grains (Weinberger et al. 2002, Sch$\ddot{u}$tz et al. 2004). Grain growth and crystallization have similar effects on the 10 $\micron$ silicate feature (Kessler-Silacci et al. 2006). As grains grow from sub-micron sizes to several microns, the 10 $\micron$ feature becomes weaker and less peaked. 2M1207 shows weak dust processing signatures compared to other TWA members. The presence of grain growth and settling in other brown dwarfs (e.g., Apai et al. 2005) does not support different dust processing mechanisms for stellar and sub-stellar objects. However, if mechanisms such as aggregate fragmentation and/or turbulent mixing occur along with grain growth in the disk, then the time scales over which grain growth and dust settling takes place may be prolonged. Fragmentation results in replenishment of small grains throughout the disk, while turbulent mixing can increase the time scale over which grains settle, as it can bring both large and small grains back to the disk surface. Grain growth models by Dullemond \& Dominik (2004) indicate that growth to meter sizes is rapid ($\sim 10^{3}$ years at 1 AU and $10^{5}$ years at 30 AU). On the other hand, fragmentation of grains allows for a semi-stationary state to be reached after about $10^{4}$ years, for grain sizes below $\sim$1 cm, which may last for several million years. Thus an equilibrium between grain growth and small grain replenishment rates may explain the presence of ISM-type grains in the upper disk layers of 2M1207 at an age of $\sim$10 Myr. Due to the regeneration of small grains on the disk surface, Kessler-Silacci et al. (2006) could not find a correlation between the strength of the 10 $\micron$ feature and the age or disk evolutionary state, which is consistent with our finding of weaker dust processing in 2M1207 compared to other TWA members. However, a detailed study of the N-band spectrum for 2M1207 would make it possible to decompose the 10 $\micron$ feature into distinct dust species and confirm our results. 

The above analysis of grain sizes is based on the 10 $\micron$ silicate feature. The flat slopes of the SED at millimeter wavelengths can correctly indicate the presence of grains larger than those in the ISM.  As shown in Fig. 3a, using sub-micron sized particles in the disk midplane provides a good fit to our observations up to 24 $\micron$. This results in steeper slopes at sub-millimeter wavelengths. Lommen et al. (2006) have found a correlation between the peak 10 $\micron$ flux and the millimeter slope for T Tauri disks, indicating that grain growth occurs in the outer disk and in the surface layers of the inner disk simultaneously. They have explained that when aggregates are fragmented in collisions, which must take place in order to preserve the small grain population in the upper disk layers to produce the 10 $\micron$ feature, the size of the fragments increases as the aggregate size increases. That is, larger particles are produced due to fragmentation of larger aggregates. For the case of TW Hya, the outer disk is optically thick and requires $\sim$1 cm sized particles to fit the sub-millimeter and millimeter slopes, while the inner disk region of $\la$4 AU is optically thin, and requires $\sim$1 $\micron$ sized grains to produce the silicate feature (Calvet et al. 2002). The presence of sub-micron sized particles in the upper layers of 2M1207 disk, as indicated by model fits, then suggests that the largest particles in the disk have not yet grown to sizes of millimeter or larger, indicating steeper slope in the millimeter regime. 

Fig. 4a shows the SEDs for mass accretion rates between $10^{-9}$ and $10^{-12} M_{\sun}$ $yr^{-1}$. Higher accretion rates result in larger scale heights in the optical and far-IR, but lower at near-IR wavelengths. Typical $\dot{M}$ values for young low mass stars and brown dwarfs are in the range of $10^{-10}$ and $10^{-12}$ $M_{\sun}$ $yr^{-1}$ (Muzerolle et al. 2005). Mohanty et al. (2003) have confirmed 2M1207 to be a CTTS-like accretor on the basis of the detection of He I and upper Balmer lines and the condition that accretors should display broad asymmetric H$\alpha$ emission. Scholz et al. (2005) have found that the accretion rate varies by a factor of 5-10 between $10^{-10.1\pm 0.7}$ and $10^{-10.8\pm 0.5} M_{\sun}$ $yr^{-1}$, on a timescale of 6 weeks.  Fig. 4a shows that while the SED for $10^{-9} M_{\sun}$ $yr^{-1}$ is clearly not a good fit, $\dot{M}$ of $10^{-10} M_{\sun}$ $yr^{-1}$ indicates excess emission in the optical and UV. The hot continuum emission seen in the optical and the ultraviolet wavelengths is produced by the accretion energy dissipated when the hot gas shocks at the
stellar surface (Hartmann 2000). We have included in Fig. 4a the photographic magnitudes B and R from the SuperCOSMOS Sky Survey (magnitudes obtained from Scholz, R. -D. et al. 2005). Though the errors for these magnitudes are large, including them rules out the model fit for $10^{-10} M_{\sun}$ $yr^{-1}$. On the other hand, model SEDs for $\dot{M}$ of $10^{-11}$and $10^{-12} M_{\sun}$ $yr^{-1}$ provide a good fit. Thus with the available observations, our model fits are consistent with the lower value of accretion rate of $\sim$$10^{-10.8} M_{\sun}$ $yr^{-1}$ reported by Scholz et al. However, the presence of C IV lines in emission in the UV spectrum (Gizis et al. 2005), that are formed when hot gas of $\sim10^{5}$ K  falls along an accretion column onto the substellar surface, and the strong evidence of funneled rather than spherical accretion (Scholz et al. 2005), indicates the presence of continuum excess emission in the optical and UV bands for 2M1207, which needs to be confirmed with future observations.

Fig. 4b shows the separate contributions of the inner rim (wall) at the dust sublimation radius (discussed below), the disk truncated at this radius, the stellar photosphere and the scattered flux, for a disk of mass $10^{-5}$$M_{\sun}$, inclination of 63$\degr$ and $\dot{M}$ of $10^{-11}$$M_{\sun}$ $yr^{-1}$. The disk models include the emission from the inner rim. In order to determine its separate contribution, we have calculated the inner rim fluxes using equations in Dullemond et al. (2001). As discussed above, contribution from the scattered light increases as the disk becomes more inclined, such that it can account for up to 90\% of the K-band flux for an edge-on disk (Walker et al. 2004). The inner rim is a major contributor to fluxes shortward of 10 $\micron$, and dominates the excess seen above the photosphere at shorter wavelengths. The flux at 10 $\micron$ is a sum of emission from the disk and the inner rim. The outer regions of the disk contribute more flux at longer wavelengths. The fractional disk luminosity for 2M1207 is found to be 0.03. This is higher than TWA 7 and 13 for which cool debris disks have been detected at 70 $\micron$ (Low et al. 2005). Other prominent stellar members in TWA like TW Hya, Hen 3-600 and HD98800B have larger $L_{IR}/L_{*}$ of the order of 0.2 due to their strong excesses at mid- and far-IR wavelengths.

The inner disk radius, $R_{in}$, was set to 1$R_{sub}$, where $R_{sub}$ is the dust sublimation radius and varies with the stellar radius and temperature, $R_{sub} = R_{*} (T_{sub}/T_{*})^{-2.085}$ (Whitney et al. 2003). $T_{sub}$ is the dust sublimation temperature and was set to 1600K. For 2M1207, 1$R_{sub} \sim$ 3$R_{*}$. Fig. 4c shows the variations in the model SEDs with increasing $R_{in}$. We have obtained good fits for $R_{in}$ between 1 and 3$R_{sub}$, with a peak at 1$R_{sub}$ (based on a reduced $\chi^{2}$ comparison). Increasing the inner radius to 5 or 7$R_{sub}$ results in higher fluxes near the 10 $\micron$ silicate band and at longer wavelengths. Inner holes of a few sub-stellar radii ($\sim$3 to 7$R_{*}$) are found to be common around brown dwarfs (e.g. Allers et al. 2006; Mohanty et al. 2004). On the basis of a lack of a K-$L^{\prime}$ excess, Jayawardhana et al. (2003) suggested an inner hole in 2M1207 disk. However, we have obtained the best fit for $R_{in}$=1$R_{sub}$, which implies an absence of an inner disk hole since it would have to be larger than the dust sublimation radius. This suggests the presence of some K-$L^{\prime}$ excess. Since the disk is relatively flat, the K-$L^{\prime}$ excess will be small, similar to the observed K-[3.6] excess, even if the inner radius extends down to the dust sublimation radius. Jayawardhana et al. have used a conservative limit of K-$L^{\prime}$ $\ga$ 0.2 to define this excess, which may be the reason for their observation to be purely photospheric. Other than TW Hya, Hen 3-600, HD98800B and HR4796A, most of the TWA members do not show any excess at near- and mid-IR wavelengths, which indicates dissipation of dusty inner disks on a time scale of $\la$10 Myr (Low et al. 2005; Jayawardhana et al. 1999). The lack of inner disk dissipation for 2M1207 could be explained by a possible non-coevality in the TWA. There are results suggesting two distinct populations in TWA, as evidenced by a bimodal distribution in the rotation periods (Lawson \& Crause 2005) and the presence of warm (T$\ga$ 100 K) dust (Low et al. 2005; Weinberger et al. 2004). Low et al. found negligible amounts of warm dust around 20 out of their 24 TWA targets, while the other four (mentioned above) display strong excess emission at 24 $\micron$. Lawson \& Crause have found the median rotation period for the TWA 1-13 group led by TW Hya to be 4.7 d, while that for the stars in the TWA 14-19 group to be 0.7 d. These authors have suggested that the former group might be younger than the latter by $\sim$8 Myr (8-10 Myr versus $\sim$17 Myr). An upper period limit of 25 hr for 2M1207 (Scholz et al. 2005) would then place it closer to the older ($\sim$17 Myr) group. Recently, Barrado y Navascu\'{e}s (2006) has estimated the age for 2M1207 to be $15^{+15}_{-10}$ Myr. Considering the large uncertainty in this estimate, if 2M1207 is indeed as young as $\sim$5 Myr, this would explain the lack of inner disk evolution in this brown dwarf disk. On the other hand, if it is as old as $\sim$30 Myr, then regeneration of small grains in the upper disk layers would be a plausible explanation for the lack of inner disk clearing, as discussed above. Given the assumption that all TWA members have similar ages, the wide variety in the SEDs is similar to what has been observed for T Tauri disks in other clusters. Lada et al. (2006) found that the IR excess due to circumstellar disks around low-mass stars in the IC 348 cluster (2-3 Myr) ranges
from optically thick flared disks to diskless photospheres. Hartmann et al. (2005) found a similar variety in the SEDs of CTTS in Taurus (0.1-2 Myr). These vary from stars with highly flared disks (e.g., DR Tau), to transitional disks (e.g., GM Aur) and stars with purely photospheric emission (e.g., IW Tau). The differences in the SEDs are attributed to the variations in the scale heights of the disks as well as the disk inclinations and disk masses, but could also be reflecting intrinsic differences in the
star systems themselves, such as different initial conditions or other intrinsic properties like the activity level of the star, or planet formation, etc.

We have varied the outer disk radius between 5 and 200 AU. Since the disk mass was fixed at $10^{-5} M_{\sun}$, smaller $R_{out}$ resulted in larger optical depth. However, the changes in the SEDs are more evident at far-IR/sub-mm wavelengths, and longer wavelength observations can correctly determine the outer disk radius. As discussed in the introduction, Bate et al. (2003) have predicted truncated disks of outer radii $<$10-20 AU, if brown dwarfs form via the ejection scenario. Though we have obtained a good fit using $R_{out}$ = 100 AU, the degeneracies in our fits do not allow us to rule out such a formation scenario for 2M1207. We are able to get similar fits by varying both disk mass and outer radius to give similar optical depths at the wavelengths being modeled. As discussed in Walker et al. (2004), it is difficult to test such predictions with SEDs alone, and high resolution imaging is required to resolve the disk via their scattered and thermal emission.

\section{Summary}

Due to the large number of free parameters and the absence of longer wavelength observations, there are degeneracies in the model fits presented here. The uncertainties in the stellar parameters could also result in changes in the disk parameters. Nevertheless, our model fits indicate that grain sizes of $a_{max} \sim$ 0.25 $\micron$ provide the best fit to the 10 $\micron$ silicate emission feature, based on a reduced $\chi^{2}$ comparison with larger grain models. In comparison with other TWA members, this suggests lesser dust processing for 2M1207, which could be explained by mechanisms such as aggregate fragmentation and/or turbulent mixing. We have obtained the best fit using an inner disk radius equal to the dust sublimation radius, indicating an absence of an inner disk hole. A flat disk of mass between $10^{-4}$ and $10^{-6}$ $M_{\sun}$, inclination between $\sim$60$\degr$ and 70$\degr$, and $\dot{M}$ of $\leq10^{-11} M_{\sun}$ $yr^{-1}$ provides a good fit to 2M1207 disk. The disk mass estimate is a lower limit; longer wavelength observations can confirm the presence of a population of much larger grains that dominate the disk mass but contribute little to the opacity at near- and mid-IR wavelengths.

\acknowledgments
We wish to thank the referee Barbara Whitney for many helpful comments and suggestions. Support for this work was provided by NASA Research Grant $\#$ NNG06GJ03G. Support for program 9841 was provided by NASA through a grant from STScI, which is operated by AURA Inc., under NASA contract NAS 5-26555. This work is based in part on observations made with the {\it Spitzer Space Telescope}, which is operated by the Jet Propulsion Laboratory, California Institute of Technology under a contract with NASA. Support for this work was provided by NASA through an award issued by JPL/Caltech. This work has made use of the SIMBAD database.

\clearpage

\begin{deluxetable}{ccrrrrrrrrcrrl}
\tabletypesize{\scriptsize}
\tablecaption{IRAC and MIPS Observations\tablenotemark{a}}
\tablewidth{0pt}
\tablehead{\colhead{{\it I}\tablenotemark{b}} & \colhead{{\it J}} & \colhead{{\it H}} & \colhead{{\it K}} &
 \colhead{3.6 $\micron$} & \colhead{4.5 $\micron$} &
\colhead{5.8 $\micron$}  & \colhead{8 $\micron$}  & \colhead{24 $\micron$}  & \colhead{3.8 $\micron$\tablenotemark{c}} & \colhead{8.7 $\micron$\tablenotemark{d}} & 
\colhead{10.4 $\micron$\tablenotemark{d}} \\
 \colhead{mJy}  & \colhead{mJy}  & \colhead{mJy}  & \colhead{mJy} & \colhead{mJy}  & \colhead{mJy}  & \colhead{mJy}  & \colhead{mJy} & \colhead{mJy}  & \colhead{mJy} &
\colhead{mJy} & \colhead{mJy}  
}
\startdata
1.13 & 10.10$\pm$0.89 & 11.35$\pm$1.0 & 11.12$\pm$0.98 & 8.49$\pm$0.32 & 7.15$\pm$0.26 &  6.36$\pm$0.06  & 5.74$\pm$0.21  & 4.32$\pm$0.03   & 7.00$\pm$0.6 & 5.60$\pm$1 & 7.50$\pm$1  \\
\enddata

\tablenotetext{a}{Riaz et al. (2006).}
\tablenotetext{b}{I and JHK magnitudes from DENIS and 2MASS, respectively.}
\tablenotetext{c}{Jayawardhana et. al. (2003).}
\tablenotetext{d}{Sterzik et. al. (2004).}

\end{deluxetable}
\clearpage

\begin{deluxetable}{cc}
\tabletypesize{\scriptsize}
\tablecaption{Stellar and disk parameters}
\tablewidth{0pt}
\tablehead{
\colhead{Parameter}  & \colhead{Value} 
}
\startdata
$M_{*}$ & 0.024 $M_{\sun}$ \\
$R_{*}$ & 0.24 $R_{\sun}$ \\
$T_{*}$ & 2550 K \\
$\beta$ & 1.05 - 1.1 \\
$M_{d}$ & $10^{-4}$ - $10^{-6}$$M_{\sun}$ \\
$\dot{M}$ & $\la10^{-11} M_{\sun}$ $yr^{-1}$ \\
$R_{min}$ & 1 $R_{sub}$ ($\sim$3 $R_{*}$) \\
{\it i} & 63$\degr$ - 69$\degr$ \\
\enddata
\end{deluxetable}

\begin{figure}
  \begin{center}
 \resizebox{100mm}{!}{\includegraphics[angle=270]{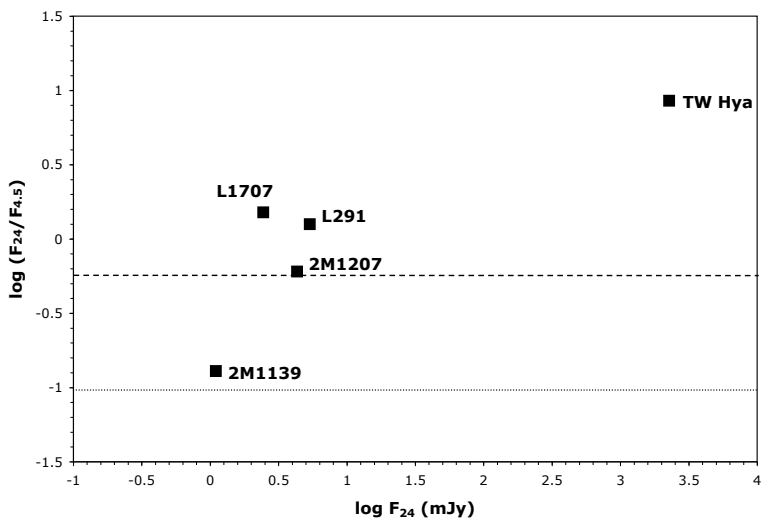}}
 \caption{Amount of flaring in 2M1207 disk. Dashed line represents a geometrically thin, optically thick flat disk with a spectral slope $\lambda F_{\lambda} \propto \lambda^{-4/3}$. Solid line represents the photosphere of a brown dwarf at $T_{eff}$=2600K. See text for more details.}
 \end{center}
\end{figure}

\begin{figure}
 \begin{center}
    \begin{tabular}{c}
      \resizebox{90mm}{!}{\includegraphics[angle=0]{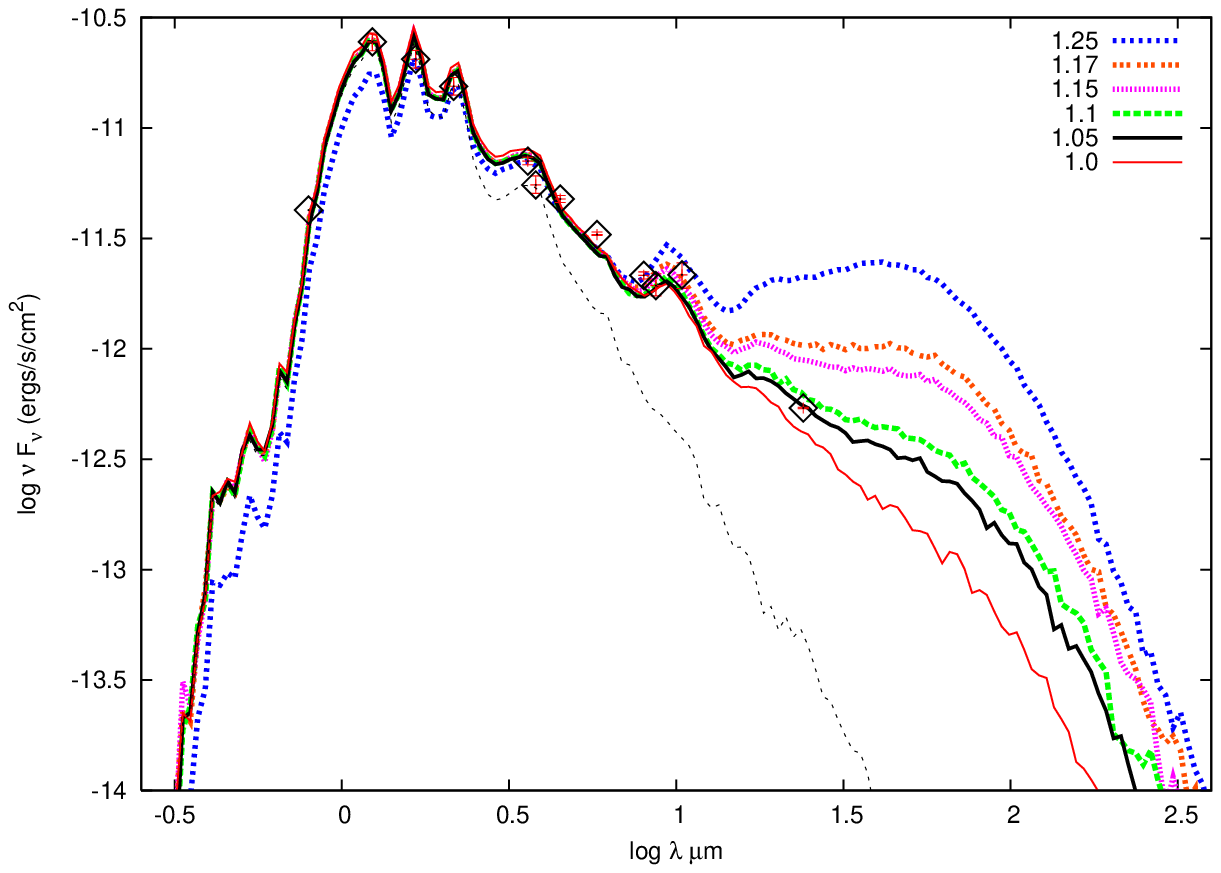}} \\
      \resizebox{90mm}{!}{\includegraphics[angle=0]{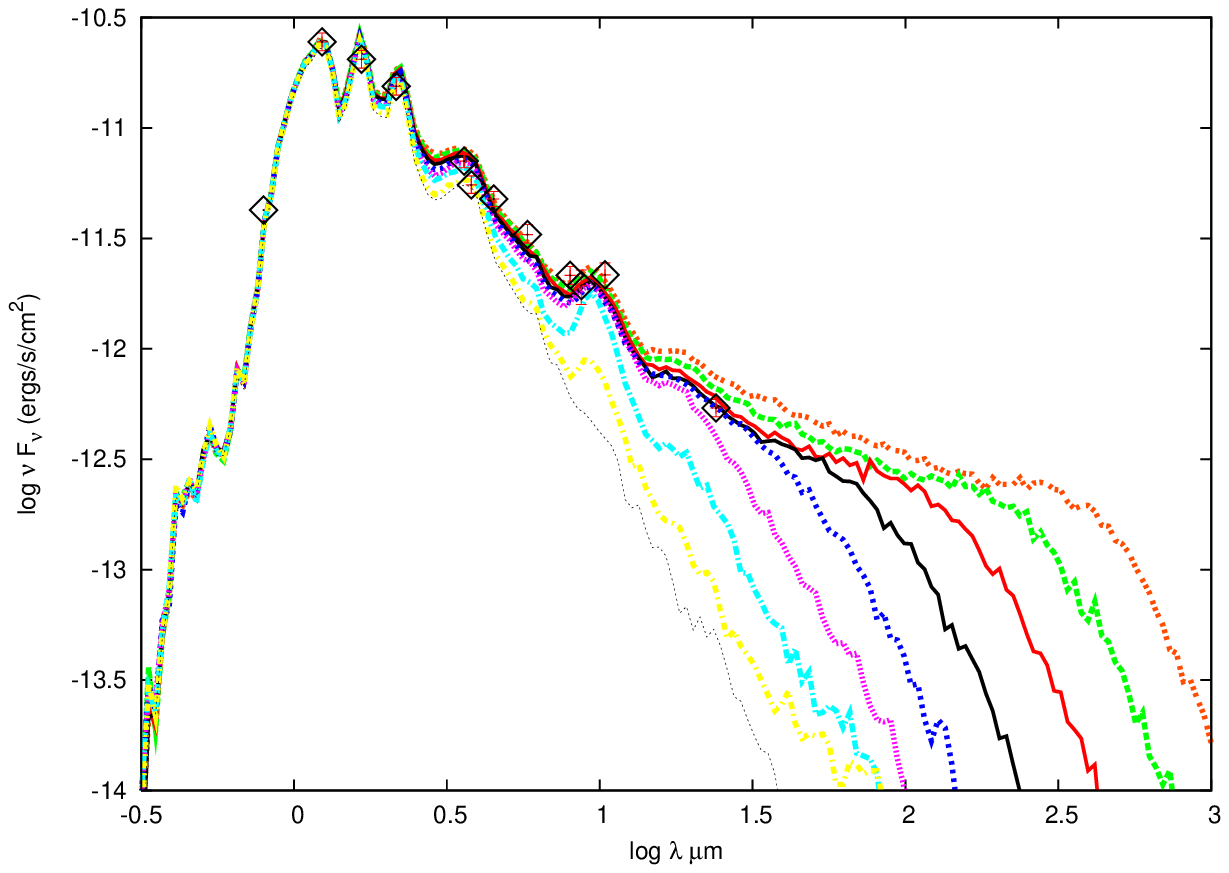}} \\
       \resizebox{90mm}{!}{\includegraphics[angle=0]{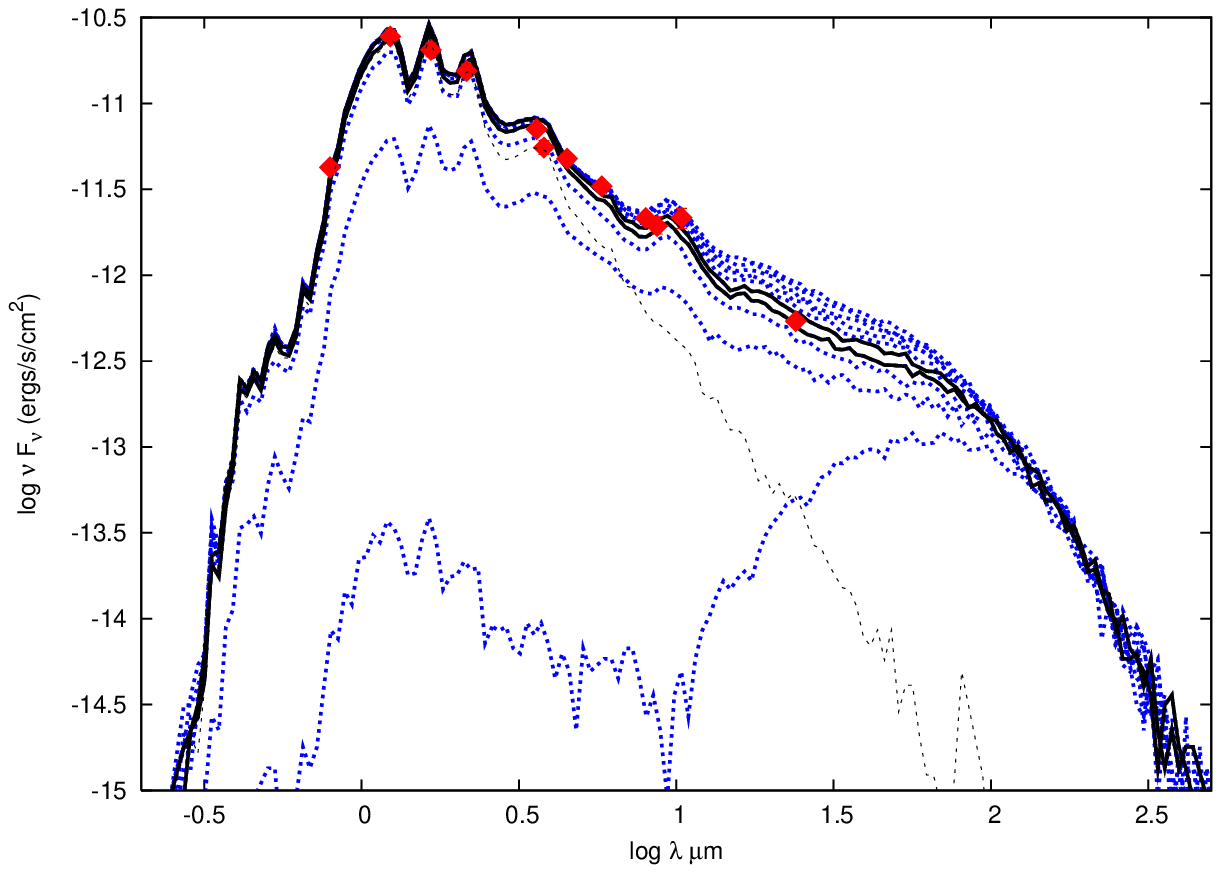}} \\
    \end{tabular}
    \caption{{\it Top}: (a) Variations in the SEDs with flaring power $\beta$. {\it Middle}: (b) Models for disk masses ranging between $10^{-2}M_{\sun}$ (top SED) and $10^{-9}M_{\sun}$ (bottom SED). {\it Bottom}: (c) SEDs for 10 viewing angles between edge-on (bottom SED) and face-on (top SED). Models for 63$\degr$ and 69$\degr$ (marked in black) provide the best fit for 2M1207. The thin black dotted line  in all figures represents the photospheric flux.}
  \end{center}
\end{figure}

\begin{figure}
 \begin{center}
    \begin{tabular}{cc}      
      \resizebox{90mm}{!}{\includegraphics[angle=0]{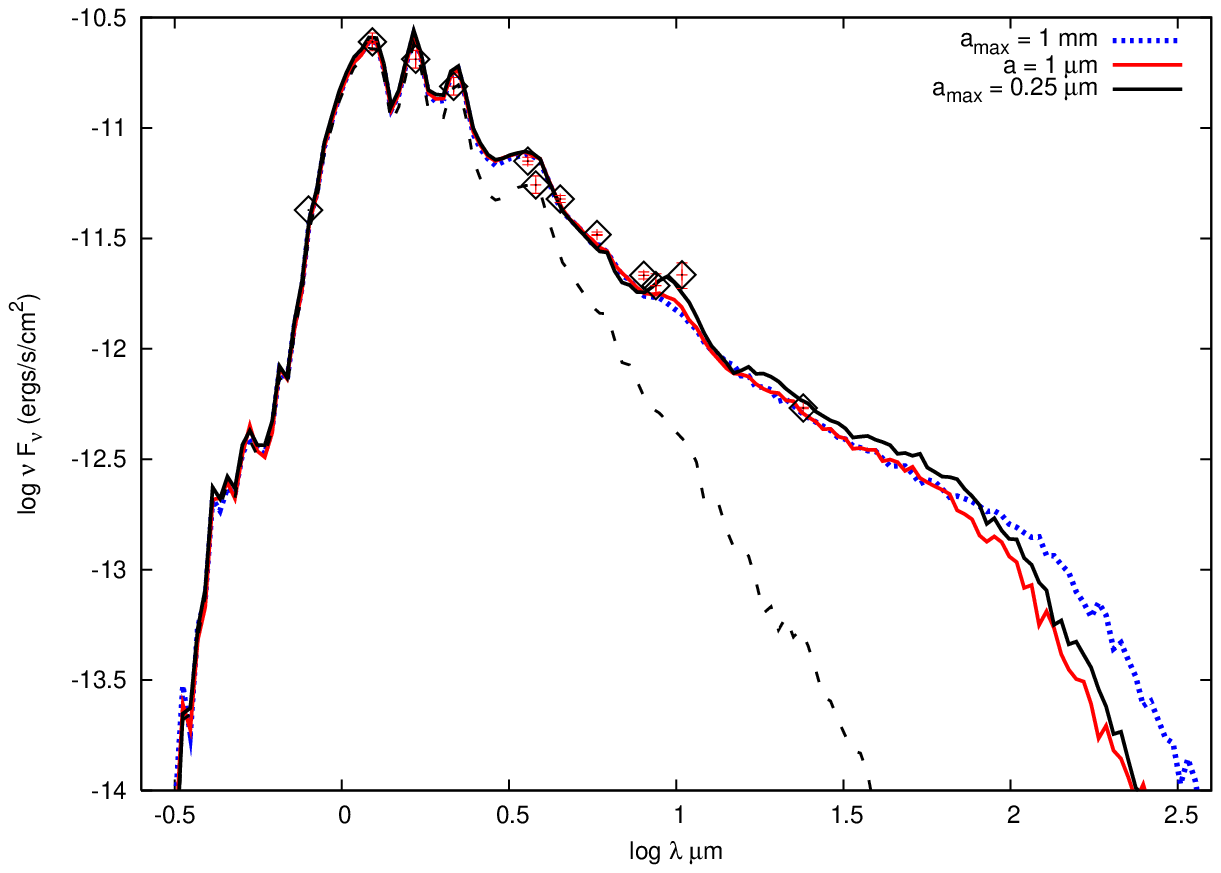}} \\
      \resizebox{90mm}{!}{\includegraphics[angle=0]{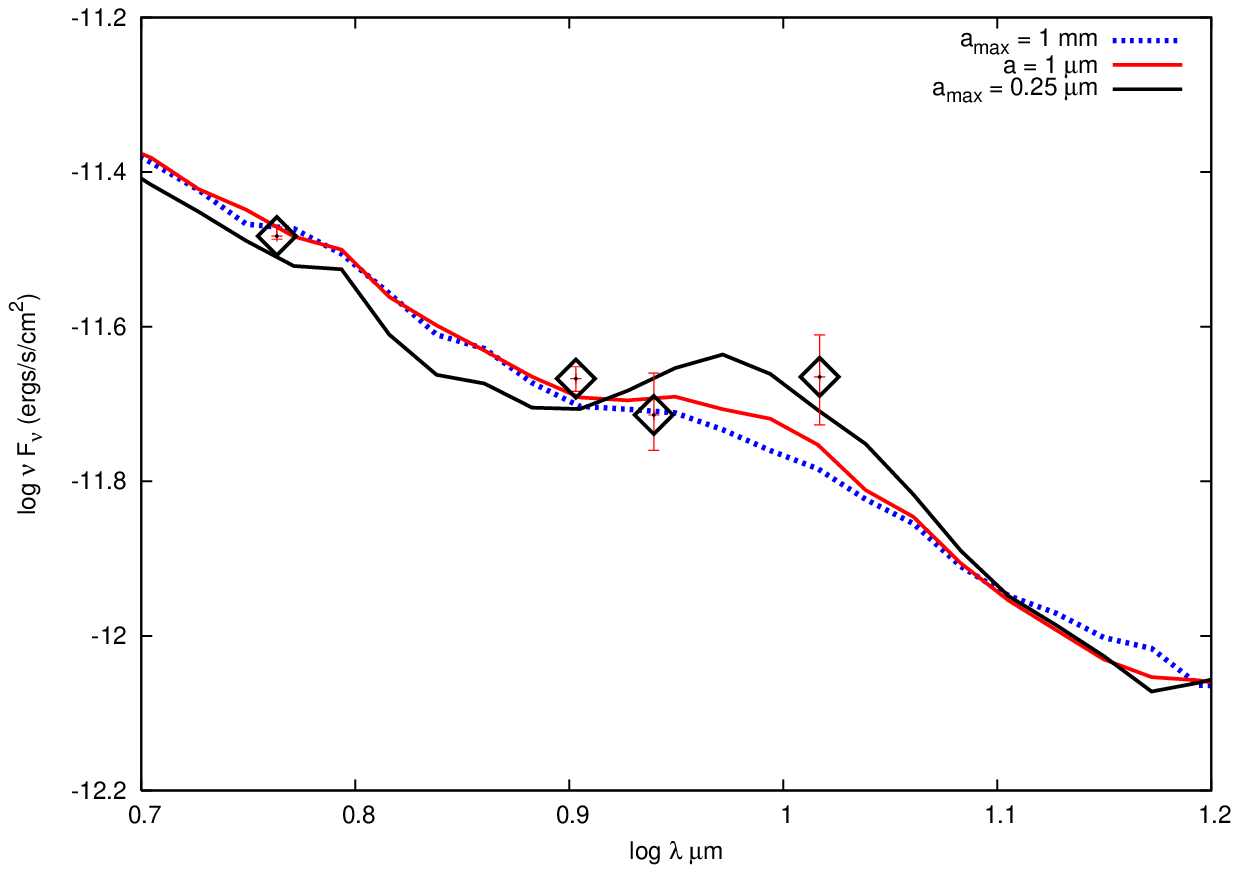}} \\
    \end{tabular}
    \caption{SEDs for three different grain models: {\it Blue} - $a_{max}$ = 1 mm, {\it red} - a = 1 $\micron$, {\it black} - $a _{max}$ = 0.25 $\micron$.  {\it Top}: (a) Variations in the SEDs are evident near 10 $\micron$ and at far-IR/sub-mm wavelengths. {\it Bottom}: (b) The 10 $\micron$ Si emission feature. Grain size of $a_{max}$ = 0.25 $\micron$ provides the best fit.}
  \end{center}
\end{figure}

\begin{figure}
 \begin{center}
    \begin{tabular}{cc}      
      \resizebox{90mm}{!}{\includegraphics[angle=0]{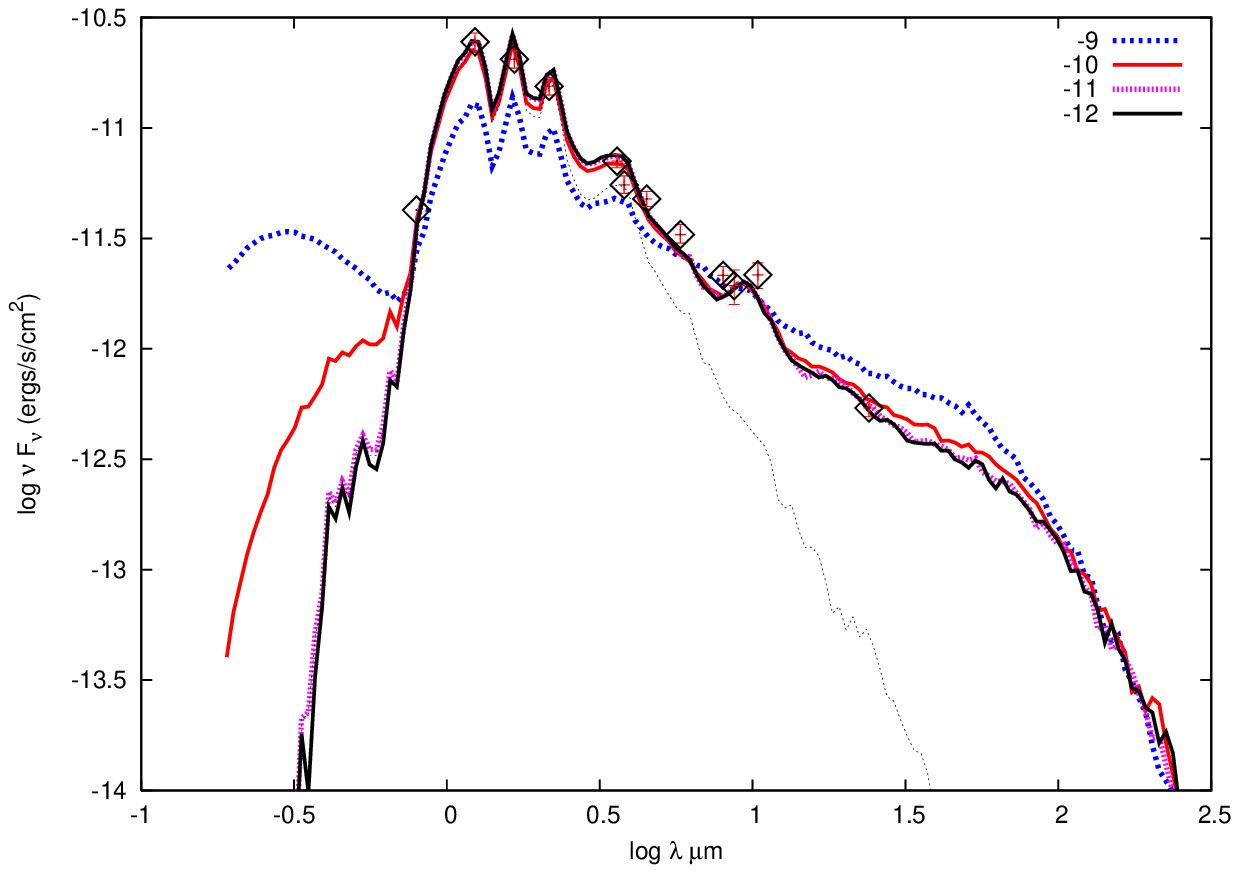}} \\
      \resizebox{90mm}{!}{\includegraphics[angle=0]{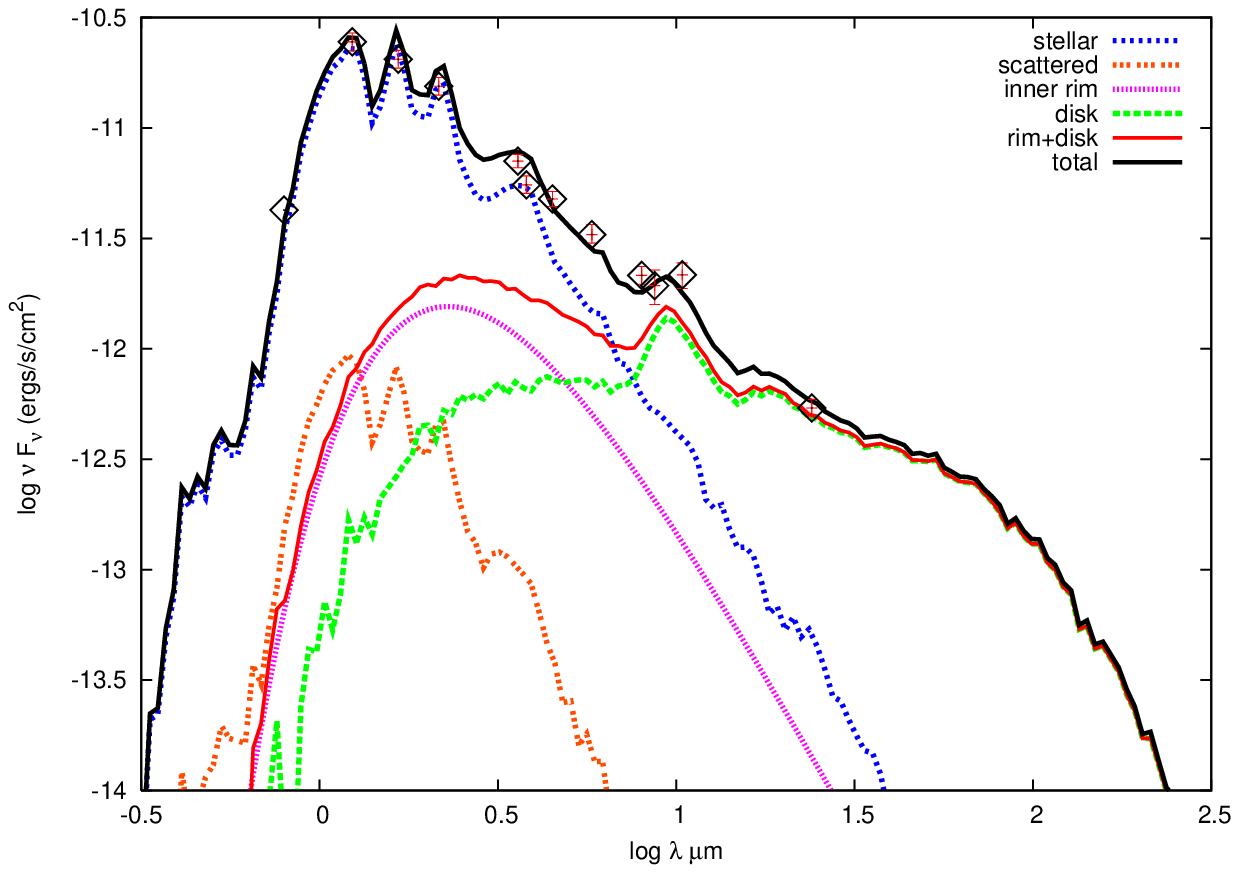}} \\
      \resizebox{90mm}{!}{\includegraphics[angle=0]{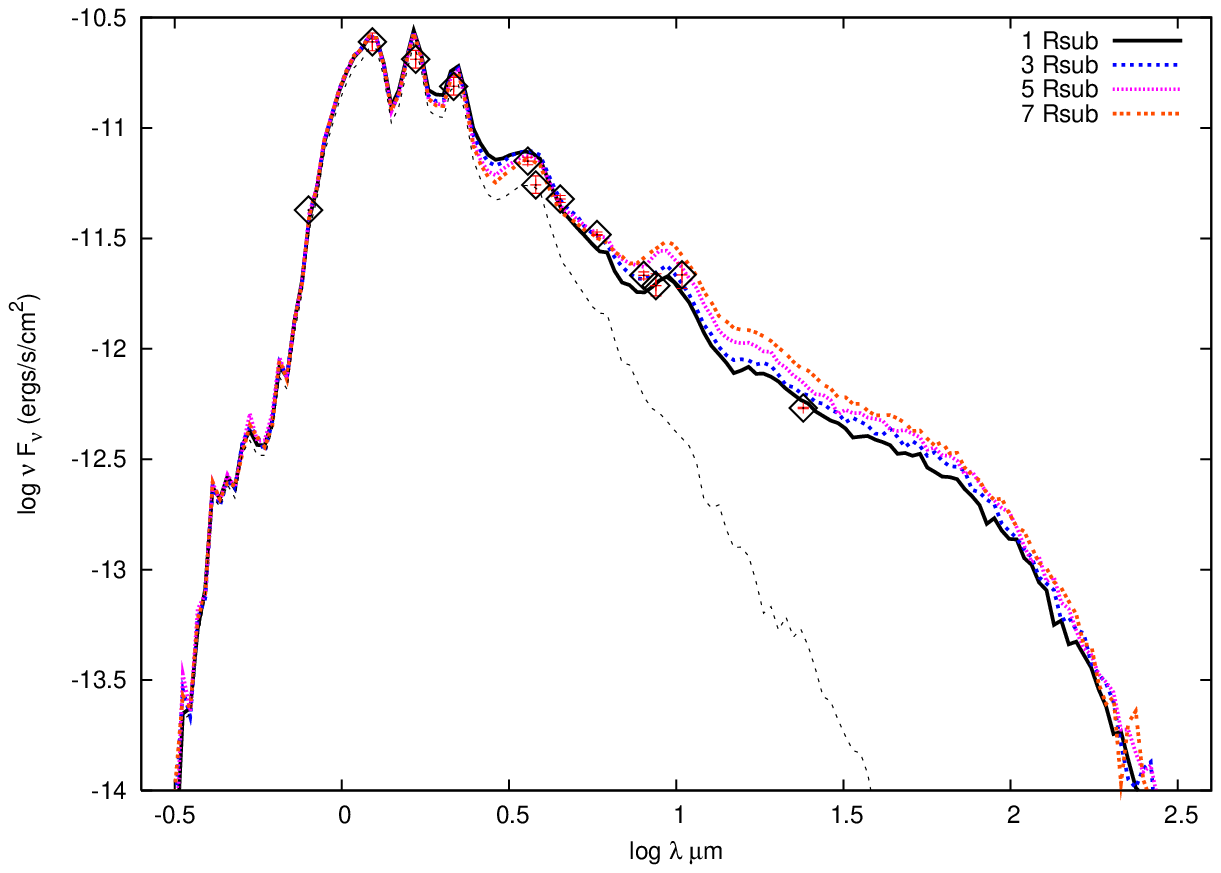}} \\      
    \end{tabular}
    \caption{{\it Top}: (a) Variations in the SEDs with disk mass accretion rates between $10^{-9}$ and $10^{-12} M_{\sun}$ $yr^{-1}$. {\it Middle}: (b) Separate contributions of the inner rim (wall) at the dust sublimation radius, the disk, the stellar photosphere and the scattered flux. {\it Bottom}: (c) Models for different inner disk radii. $R_{in}$=1$R_{sub}$ provides the best fit for 2M1207.  }
  \end{center}
\end{figure}

\end{document}